\documentclass[prb,superscriptaddress,showpacs,floatfix,twocolumn]{revtex4}
\usepackage{amssymb}
\usepackage{amsmath}
\usepackage{graphicx}

\begin{document}

\title{Solitonic Excitations in Linearly Coherent Channels of Bilayer
Quantum Hall Stripes }
\author{C. B. Doiron}
\altaffiliation[Current address: ]{Department of Physics and Astronomy, University of Basel, 4056 Basel,
Switzerland.}
\author{R. C\^{o}t\'{e}}
\email{Rene.Cote@Usherbrooke.ca}
\affiliation{D\'{e}partement de physique and RQMP, Universit\'{e} de Sherbrooke,
Sherbrooke, Qu\'{e}bec, Canada, J1K 2R1}
\author{H. A. Fertig}
\affiliation{Department of Physics, Indiana University, Bloomington, Indiana 47405}

\begin{abstract}
In some range of interlayer distances, the ground state of the
two-dimensional electron gas at filling factor $\nu =4N+1$ with $N=0,1,2,...$
is a coherent stripe phase in the Hartree-Fock approximation. This phase has
one-dimensional coherent channels that support charged excitations in the
form of pseudospin solitons. In this work, we compute the transport gap of
the coherent striped phase due to the creation of soliton-antisoliton pairs
using a supercell microscopic unrestricted Hartree-Fock approach. We study
this gap as a function of interlayer distance and tunneling amplitude. Our
calculations confirm that the soliton-antisoliton excitation energy is lower
than the corresponding Hartree-Fock electron-hole pair energy. We compare
our results with estimates of the transport gap obtained from a
field-theoretic model valid in the limit of slowly varying pseudospin
textures.
\end{abstract}

\date{\today }
\maketitle

\section{Introduction}

It is well known that the ground state of the two-dimensional electron gas
(2DEG) in single layer quantum Hall systems near half-odd integer filling
factors in Landau levels $N\geq 2$ \textit{i.e.} for $\nu =9/2,11/2,\ldots $
is a striped state responsible for a strong anisotropy in the conductivity
tensor of the 2DEG. This state was predicted on the basis of Hartree-Fock
calculations\cite{stripetheory} and has been extensively studied
experimentally.\cite{stripeexperimental}

When the interlayer distance, $d$, in a bilayer quantum Hall system at
filling factor $\nu $ is large, one expects the system to behave as two
isolated two-dimensional electron gases (2DEG) with filling factor $\nu /2$.
It is then natural to infer that the ground state of the 2DEG in a bilayer
should be a striped state at $\nu =4N+1$ at sufficiently large interlayer
distances. On the other hand, it is known that, at $\nu =4N+1$ interlayer
interactions can lead to a homogeneous ground state with spontaneous phase
coherence between the layers when the interlayer distance is comparable with
the separation between electrons in a single layer. One might then
conjecture that, as the interlayer separation is decreased, the striped
state acquires a certain degree of coherence due to the interlayer
interaction. This conjecture was first studied by Brey and Fertig\cite%
{breystripes} who showed that, as $d$ is increased from zero the bilayer
ground state goes from a uniform coherent state (UCS) at small interlayer
separations to a coherent striped phase (CSP)\ at $d\geq d_{1}$ and then
into a modulated striped state (or anisotropic Wigner crystal) at $d\geq
d_{2}$. The interlayer coherence is lost in the modulated stripe state. The
range $\left[ d_{1},d_{2}\right] $ increases with $N$.\cite{Dorra}

\begin{figure}[tbph]
\includegraphics[scale=1]{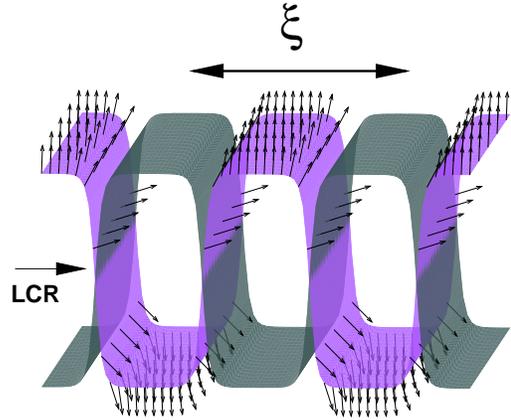}
\caption{Guiding center density in the right (dark surface) and left (light
surface) wells and pseudospin pattern in the coherent stripe phase. The arrow
indicates one linearly coherent channel (LCR). }
\label{pattern}
\end{figure}

The coherent striped phase shown in Fig. \ref{pattern} is a state where
charge density waves in the two layers are shifted by $\xi /2$ where $\xi $
is the period of the stripes in one layer. The most interesting aspect of
the CSP is that in the regions where the charge densities in both layers
\textquotedblleft overlap\textquotedblright\ (in the plane of the
two-dimensional electron gas (2DEG)), the electrons are effectively in a
linear superposition of states of the form $\left\vert \psi \right\rangle
=\left( \left\vert R\right\rangle +\left\vert L\right\rangle \right) /\sqrt{2%
}$ where $R,L$ indicates the right and left wells. The interlayer coherence
is then maintained but only along linearly coherent regions (LCR's) whose
width decreases as $d$ increases. The CSP is most easily represented in the
pseudospin language where an up (down) pseudospin is associated with the
right (left) well. The CSP is a pseudospin density wave where the
pseudospins oscillates in the $xz$ plane and the LCR's are the
one-dimensional regions where the pseudospins lie along the $x$ direction in
the $xy$ plane. 

In a previous work\cite{cotecspmodes}, we have computed the collective
excitations of the CSP and showed that the low-energy modes of this phase
could be described by an effective pseudospin wave hamiltonian. We have also
shown\cite{coteparallel} that the application of a parallel magnetic field
gives rise to a very rich phase diagram for the 2DEG involving
commensurate-incommensurate transitions with distinctive signatures in the
collective excitations and tunneling $I-V$. A very exhaustive study of the
phase diagram of the 2DEG in the presence of a parallel magnetic field, in
higher Landau levels, has also been published by Daw-Wei Wang et al.\cite%
{demlerlongarticle,demlerarticlecourt}.

\begin{figure}[tbph]
\includegraphics[scale=1]{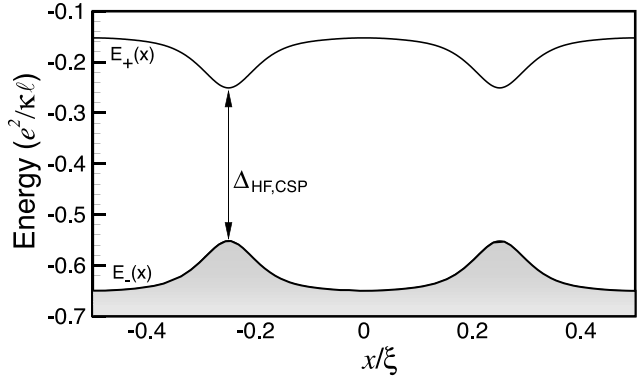}
\caption{Band structure of the coherent stripe phase. The greyed states
represent filled states at $\protect\nu=4N+1$. The Hartree-Fock gap is also
indicated. It corresponds to the excitation of an electron-hole pair in one
of the linearly coherent channels.}
\label{bandestructure}
\end{figure}

The band structure of the CSP is shown in Fig. \ref{bandestructure}. In the
Hartree-Fock approximation, the energy gap of this system corresponds to the
excitation of an electron-hole pair in a coherent channel (a pseudospin flip
in the $xy$ plane) and is finite if the tunneling parameter $t\neq 0.$ An
estimate of this gap, taking into account some quantum fluctuations, has
been done by E. Papa \textit{et al}.\cite{Papamacdo}. However, Brey and
Fertig\cite{breystripes} pointed out that, in analogy with spin (pseudospin)
skyrmion excitation in single (double) layer quantum Hall systems at $\nu =1$%
, the lowest-energy charged excitation should be a pseudospin soliton (or
antisoliton) in a coherent channel and the gap should be given by the energy
required to create a soliton-antisoliton pair. A pseudospin soliton of
charge $q=e$ corresponds to a $2\pi $ rotation of the pseudospin in the $xy$
plane. As for skyrmions or bimerons, the size of these solitons is
determined by a competition between tunneling energy (which favors small
solitons) and interwell exchange energy and Coulomb interaction which favors
slowly varying pseudospin textures (large solitons).

In this work, we compute the energy gap of the CSP\ due to the excitation of
a soliton-antisoliton pair as a function of tunneling and interlayer
distance. We use a supercell microscopic unrestricted Hartree-Fock approach
to extract the energy of a single soliton from that of a crystal of solitons
localized in the LCR's at filling factor $\nu =4N+1+\Delta \nu $. Our
calculation shows that a soliton-antisoliton pair has a lower energy than
the electron-hole pair so that these topological excitations will be
important in determining the transport properties of the CSP. For
completeness, we also compute the energy gap of the CSP\ using a simple
field-theoretic model based on the sine-Gordon Hamiltonian where an exact
solution for the pseudospin soliton can be obtained. This model does not
contain all the terms included in the microscopic approach, but, for slowly
varying pseudospin textures, it should give a fair estimate of the energy
gap. We actually improve on this model by taking into account that the
channels have a width that depends on the interlayer distance $d$ and also
by taking into account the interaction of the pseudospins in different
channels and the Coulomb interaction between different portions of the
topological charge densities.

The paper is organized as follows. In Sec. II, we describe the phase diagram
of the 2DEG in the bilayer system at filling factors $\nu =4N+1$ and $\nu
=4N+1+\Delta \nu $ and define the domain of existence of the soliton crystal
from which we want to compute the soliton energy. In Sec. III, we introduce
the simple field-theoretic model and the exact solution for the pseudospin
sine-Gordon solution. Section IV\ discusses the supercell method that we use
to extract the energy of a single soliton from that of a crystal of
solitons. The removal of the soliton-soliton energy is discussed in Sec. V.
Section VI discusses our numerical results. We conclude in Sec. VII. Details
of the derivation of the microscopic expression for the parameters of the
field-theoretic model are given in the appendix.

\section{Phase diagram of the 2DEG around $\protect\nu =4N+1$}

In this section, we review the phase diagram of the 2DEG at filling factor $%
\nu =4N+1$ where the coherent striped state is found and at filling factors
slightly above $\nu =4N+1$ in order to find the range of interlayer
distances where a crystal of solitons localized in the LCR's is stable. We
need the energy of this soliton lattice in order to compute the gap energy
as we explained in the introduction. To establish the phase diagram, we
compute the energy of different electronic phases in the Hartree-Fock
approximation in order to find the one that minimizes the total energy at a
given value of $\nu ,d,$ and $t$. The order parameters for the different
phases are the expectation values of the density operator projected onto the
Landau level $N$ of the partially filled Landau level (the guiding center
density), \textit{i.e.}, 
\begin{eqnarray}
\left\langle \rho _{N}^{i,j}(\mathbf{q})\right\rangle &=&\frac{1}{N_{\phi }}%
\sum_{X,X^{\prime }}e^{-iq_{x}(X+X^{\prime })/2}\delta _{X,X^{\prime
}-q_{y}\ell ^{2}} \\
&&\times \left\langle c_{X,i,N}^{\dagger }c_{X^{\prime },j,N}\right\rangle ,
\notag
\end{eqnarray}%
where $i,j$ are layer indices and $X,~X^{\prime }$ are guiding center
coordinates\cite{cotemethodewc}. We make the usual approximation of assuming
that the filled levels are inert. We also neglect Landau level mixing and
assume that the electron gas in the partially filled level is fully spin
polarized. In a crystal phase, $\left\langle \rho _{N}^{i,j}(\mathbf{q}%
)\right\rangle $ is non zero only for $\mathbf{q=G}$ where $\mathbf{G}$ is a
reciprocal lattice vector of the crystal. Defining the Hartree and Fock
interactions

\begin{widetext} 
\begin{equation}
H_{i,j}(N,M;\mathbf{q})=\frac{1}{q\ell }\Lambda _{i,j}(\mathbf{q}%
)e^{-q^{2}\ell ^{2}/2}L_{N}^{0}\left( \frac{q^{2}\ell ^{2}}{2}\right)
L_{M}^{0}\left( \frac{q^{2}\ell ^{2}}{2}\right) ,
\end{equation}%
and%
\begin{equation}
X_{i,j}(N,M;\mathbf{q})=\frac{\left[ \min \left( M,N\right) \right] !}{\left[
\max \left( M,N\right) \right] !}\int_{0}^{\infty }dy\left( \frac{y^{2}}{2}%
\right) ^{\left\vert N-M\right\vert }e^{-y^{2}/2}\left[ L_{\min \left(
N,M\right) }^{\left\vert N-M\right\vert }\left( \frac{y^{2}}{2}\right) %
\right] ^{2}\Lambda _{i,j}\left( \frac{y}{\ell }\right) J_{0}\left( q\ell
y\right) ,
\end{equation}%
\end{widetext}where $L_{N}^{M}\left( x\right) $ is a generalized Laguerre
polynomial, $J_{0}\left( x\right) $ is the zeroth-order Bessel function of
the first kind and the form factor%
\begin{equation}
\Lambda_{i,j} = 
\begin{cases}
1, & \text{if $i=j$}, \\ 
e^{-qd}, & \text{if $i \ne j$},%
\end{cases}%
\end{equation}%
the Hartree-Fock energy per electron at total filling factor $\nu =4N+%
\widetilde{\nu }$ can be written as%
\begin{equation}
\frac{E}{N_{e}}=\varepsilon \left( \frac{e^{2}}{\kappa \ell }\right) ,
\end{equation}%
with

\begin{eqnarray}
\varepsilon &=&-\frac{2\widetilde{t}}{\nu }\mathrm{Re}\left[ \langle \rho
_{N}^{R,L}(0)\rangle \right]  \label{2_2p} \\
&&+\frac{1}{2\nu }\sum_{i,j}\sum_{\mathbf{G}\neq 0}H_{i,j}\left( N,N,\mathbf{%
G}\right) \left\langle \rho _{N}^{i,i}\left( -\mathbf{G}\right)
\right\rangle \left\langle \rho _{N}^{j,j}\left( \mathbf{G}\right)
\right\rangle  \notag \\
&&-\frac{1}{2\nu }\sum_{i,j}\sum_{\mathbf{G}}X_{i,j}\left( N,N,\mathbf{G}%
\right) \left\langle \rho _{N}^{i,j}\left( -\mathbf{G}\right) \right\rangle
\left\langle \rho _{N}^{j,i}\left( \mathbf{G}\right) \right\rangle  \notag \\
&&-\frac{2}{\nu }\sum_{n<N}\sum_{n^{\prime }<N}X_{R,R}\left( n,n^{\prime
},0\right)  \notag \\
&&-\frac{1}{\nu }\sum_{n<N}X_{i,i}\left( n,N,0\right) \widetilde{\nu }. 
\notag
\end{eqnarray}%
In this last equation, $N_{e}$ is the total number of electrons in the 2DEG, 
$\widetilde{t}$ is the tunneling strength (in units of $\left( e^{2}/\kappa
\ell \right) $, with $\kappa $ the dielectric constant of the host material
and $\ell =\sqrt{\hbar c/eB}$ the magnetic length).

The last two terms in Eq. (\ref{2_2p}) give the interaction between
electrons in the filled levels and between electrons in the filled levels
and electrons in the partially filled level $N$. As we will show later, the
filled levels contribute to the quasiparticle energies, but not to the
charge gap.

The set of $\langle \rho _{N}^{i,j}(\mathbf{G})\rangle$'s corresponding to
one particular electronic phase is found by solving the equation of motion
for the one-particle Green's function in the Hartree-Fock approximation. The
method is described in detail in Ref.~\onlinecite{cotemethodewc}.

The band structure of the CSP contains two bands $E_{\pm }\left( X\right) $,
as shown in Fig. \ref{bandestructure}. At exactly $\nu =4N+1$, the
lowest-energy band is completely filled and the system is gapped even in the
absence of tunneling. In fact, in the uniform coherent state that occurs for
values of $d$ for which stripe ordering had not set in, the band structure
consists of two straight lines separated by a gap $\Delta _{UCS}=\left( 2%
\widetilde{t}+2X_{R,L}(N,N;0)\right) \left( e^{2}/\kappa \ell \right) $ with 
$\Delta_{UCS} \rightarrow 2\widetilde{t}$ as $d\rightarrow \infty .$ In the CSP,
the energy bands are periodically modulated in space with the maxima
(minima) of the valence (conduction) band at the locations of the LCR's. At
the Hartree-Fock level, the energy gap is the energy needed to excite an
electron from a maximum of the valence band to a minimum of the conduction
band. This excitation corresponds to a single spin flip localized in one
LCR. The decrease in the HF\ gap in the CSP is due not so much to the
reduction of $X_{R,L}(N,N;0)$ with $d$ as to the increase in intralayer
correlations that increases the with of the modulations in $E_{\pm }\left(
X\right)$. As $d$ increases, the charge modulations get sharper up to the
point where the stripes become square waves at very large $d$.
Correspondingly, the width of the LCR's decreases with $d$ since interwell
coherence and charge modulation compete with each other.

In analogy with the excitations of skyrmions in single quantum well and
bimerons in bilayer systems at $\nu =1$, Brey and Fertig\cite{breystripes}
noted that a lower-energy excitation could be achieved by exciting a
pseudospin soliton in the LCR instead of a simple electron-hole pair. The
pseudospin soliton corresponds to a $2\pi $ rotation of the pseudospin in
one LCR. A slowly varying pseudospin configuration like that in a soliton
has lower exchange energy than a single pseudospin flip but the cost in
tunneling energy is increased. As for skyrmions or bimerons, an optimal size
for the soliton is obtained at given values of $\nu ,d$ and $t$. The energy
cost for this optimal soliton should be compared with the Hartree-Fock
electron-hole pair excitation to determine whether or not these topological 
excitations are energetically favorable.

In a quantum Hall system, the relation between the charge density of the
solitons and their pseudospin texture (at $\widetilde{\nu }=1$) is given by
the Pontryagian density\cite{macdobible} 
\begin{equation}
\delta \left\langle \rho \left( \mathbf{r}\right) \right\rangle =\frac{1}{%
8\pi N_{\phi }}\varepsilon _{abc}S_{a}\left( \mathbf{r}\right) \varepsilon
_{ij}\partial _{i}S_{b}\left( \mathbf{r}\right) \partial _{j}S_{c}\left( 
\mathbf{r}\right) ,  \label{topolo}
\end{equation}%
where $\varepsilon _{ij}$ and $\varepsilon _{abc}$ are antisymmetric tensors
and $\mathbf{S}\left( \mathbf{r}\right) $ is a classical field with unit
modulus representing the pseudospins and $\delta \left\langle \rho \left( 
\mathbf{r}\right) \right\rangle $ is the guiding-center density. If we write
a general solution as

\begin{eqnarray}
S_{x}\left( \mathbf{r}\right) &=&\sin \theta \left( \mathbf{r}\right) \cos
\varphi \left( \mathbf{r}\right) , \\
S_{y}\left( \mathbf{r}\right) &=&\sin \theta \left( \mathbf{r}\right) \sin
\varphi \left( \mathbf{r}\right) , \\
S_{z}\left( \mathbf{r}\right) &=&\cos \theta \left( \mathbf{r}\right) ,
\end{eqnarray}%
then the induced density takes the simple form 
\begin{equation}
\delta \rho \left( \mathbf{r}\right) =\frac{1}{4\pi N_{\phi }}\sin \theta
\left( \mathbf{r}\right) \left[ \nabla \varphi \left( \mathbf{r}\right)
\times \nabla \theta \left( \mathbf{r}\right) \right] \cdot \widehat{\mathbf{%
z}}.  \label{pontry}
\end{equation}%
In a LCR, the polar angle of the pseudospins $\theta =\pi /2$. If a soliton
is present in this LCR, then $\varphi \left( \mathbf{r}\right) $ rotates by $%
\pm 2\pi $ along the channel (oriented in the $y$ direction). As discussed
below, this is a generalization of a soliton in the sine-Gordon model\cite%
{rajaraman}. We also have that, in the CSP, $\nabla \theta \left( \mathbf{r}%
\right) \neq 0$ in the LCR's and so the solitons carry a charge by virtue of
Eq. (\ref{pontry}).

In the case where pseudospin solitons are the lowest-energy excitations of
the CSP, we expect that the ground state at $\nu =4N+\widetilde{\nu }$ will
be a crystal of solitons localized in the LCR's. Table I shows that the
range of interlayer distances where the CSP is the system's ground state at $%
\nu =4N+1$ increases with the Landau level index. In this work, we choose to
study the phase diagram in Landau level $N=2$. We show in Fig. \ref%
{gsenergyp} the energy per electron for different electronic phases in $N=2$
as a function of interlayer distances and for three values of the tunneling
parameter $\widetilde{t}=0,0.01$ and $0.06.$ The filling factor is $\nu =9.2$%
. The contribution from the filled levels is not included in this
calculation since it depends only on $\nu $ and is thus the same for all
phases. At small interlayer distances, where the ground state at $\nu =9$ is
a UCS, the ground state at $\nu =9.2$ is a one-component hexagonal Wigner
crystal (HWC). In this phase, a crystal of electrons of pseudospin $%
S_{x}=-1/2$ and filling $\widetilde{\nu }=0.2$ sits on top of a liquid of
pseudospins $S_{x}=+1/2$ and filling $9.0$. There is no pseudospin texture
in that state and, in particular, no bimerons in contrast with the situation
in the lowest Landau level\cite{breybimeron} where the ground state is a
crystal of bimerons. In fact, we find that bimeron excitations are not
relevant in $N=2$ even in the limit of vanishing $\widetilde{t}$. 
\begin{table}[tbp]
\begin{ruledtabular}
\begin{tabular}{clll}
Landau level & $d_{1}/\ell $ & $d_{2}/\ell $ & $D/\ell $ \\ 
\colrule0 & 1.2 & 1.65 & 0.45 \\ 
1 & 0.8 & 1.45 & 0.65 \\ 
2 & 0.6 & 1.6 & 1.00 \\ 
\end{tabular}%
\end{ruledtabular}
\caption{Critical interlayer distances $d_{1}/\ell $ and $d_{2}/\ell $ at $%
\widetilde{t}=0$ for the transition UCS-CSP and CSP-modulated striped state.
The last column gives the range of interlayer distances $D/\ell =d_{2}/\ell
-d_{1}/\ell $ for which the CSP is the ground state in Landau level $N.$}
\end{table}

\begin{figure}[tbph]
\includegraphics[scale=1]{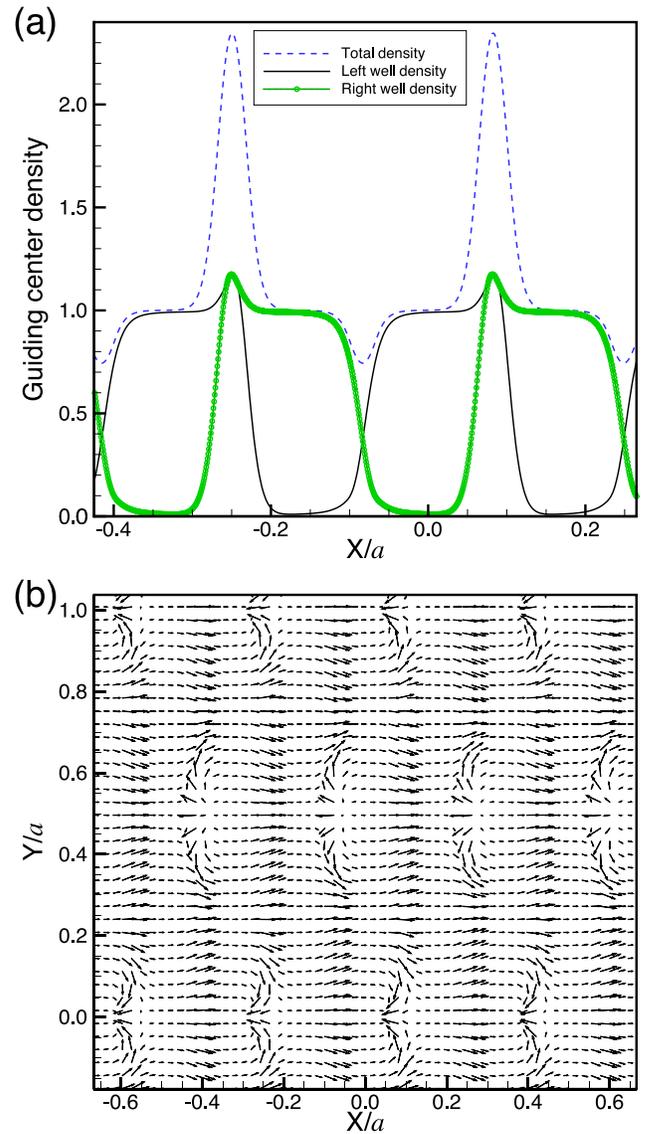}
\caption{Representation of the soliton crystal at $d/\ell =1.2$, $\widetilde{%
t}=0.01$ and $\protect\nu =9.1.$The distance between two solitons in a
channel is $a.$ (a) Guiding-center densities $\protect\rho _{RR}\left(
x,y\right) ,\protect\rho _{LL}\left( x,y\right) $ and $\protect\rho \left(
x,y\right) =\protect\rho _{RR}\left( x,y\right) +\protect\rho _{LL}\left(
x,y\right) $ at $y=0$; (b) pseudospin texture showing the solitons localized
in the channels. }
\label{solitoncristal}
\end{figure}

For interlayer distances where the CSP is found at $\nu =9$, the ground
state of the 2DEG at $\nu =9.2$ is a centered crystal of pseudospin solitons
localized in the LCR's. We note that there are many possible choices for the
lattice structure of this crystal, since solitons may or may not be present in
every LCR, depending on the commensuration of the lattice of solitons and
the underlying stripe state, and it is likely that there are phase
transitions among these different states as the filling factor is varied.
For the choice of parameters in this study, the lowest energy state has
solitons in every channel. We found however that a similar state with
solitons in every second channel but with the same filling factor has very
nearly the same energy. 
\begin{figure}[tbph]
\includegraphics[scale=1]{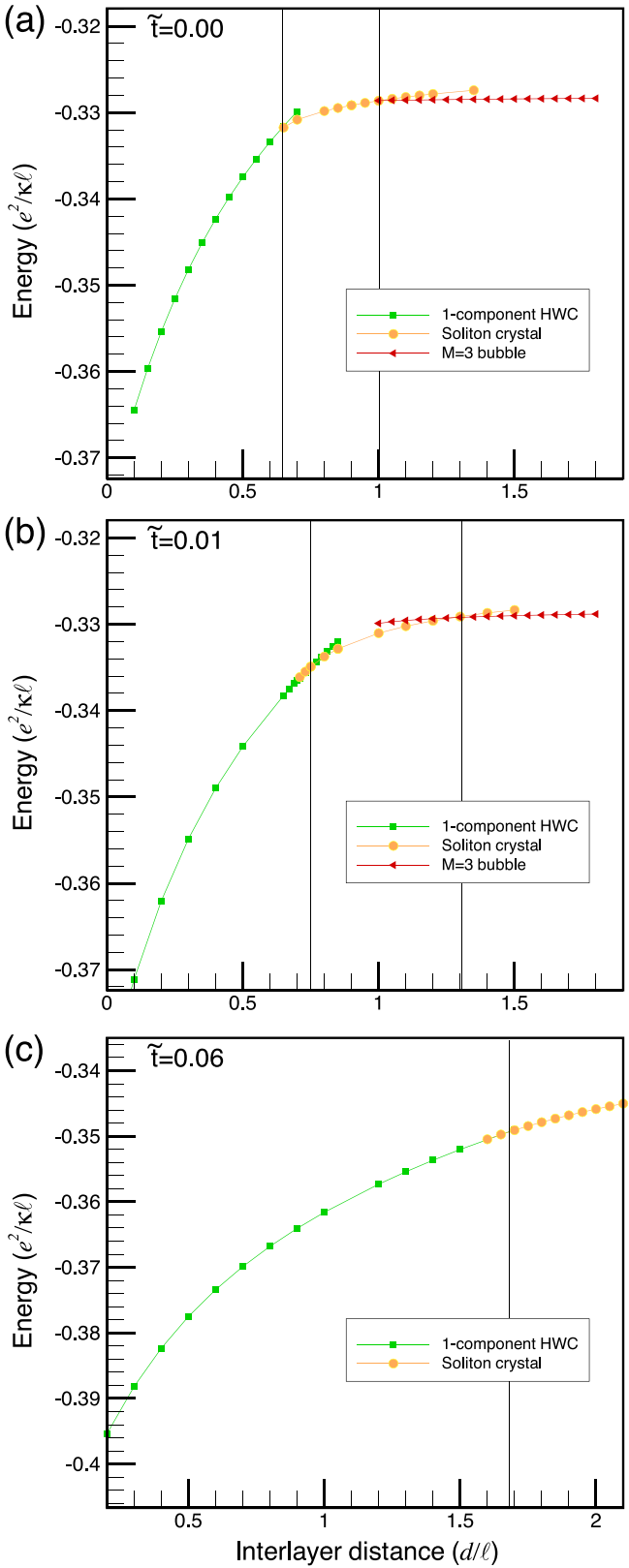}
\caption{Hartree-Fock ground state energy per electron as a function of
interlayer distances at filling factor $\protect\nu =9.2$ and for (a) $%
\widetilde{t}=0$; (b) $\widetilde{t}=0.01$; (c) $\widetilde{t}=0.06$. The
vertical lines indicate the position of the phase transitions.}
\label{gsenergyp}
\end{figure}
Figure \ref{solitoncristal} shows an example of the charge distribution as
well as the pseudospin texture associated with a centered rectangular
soliton crystal. Since the focus of this study is on the energetics of
single solitons, we will use only the structure illustrated in Fig. \ref%
{solitoncristal} for our quantitative analysis below.

At large interlayer distances, we find that the ground state of the 2DEG at $%
\nu =9.2$ is a superposition of two shifted triangular bubble crystals\cite%
{stripetheory} with partial filling $\widetilde{\nu }=0.6$ in each well.
Because $\widetilde{\nu }>0.5$ the bubbles are clusters of holes and not
electrons. We find that the number of holes per bubble is $M=3$ in agreement
with previous Hartree-Fock calculation in single quantum well systems\cite%
{cotebubble}. 

\section{Field-theoretic model}

We use two different approaches to compute the energy gap due to the
excitation of soliton-antisoliton pairs. The first one is a field-theoretic
calculation valid in the limit of slowly varying pseudospin textures. It is
explained in this section. The second one is a microscopic approach where
the energy of one soliton is computed from that of a crystal of solitons by
removing the soliton-soliton interaction. We call this method the supercell
approach. In principle, this second method is not restricted to small
gradient of the pseudospin texture and includes terms neglected in the
field-theoretic model. We expect it to be more accurate than the
field-theoretic approach.

In the field-theoretic approach, we evaluate the energy to create a
pseudospin soliton by making a long-wavelength expansion of certain terms in
the Hartree-Fock Hamiltonian. We follow the procedure developped in details
in Ref.~\onlinecite{macdobible}. To keep the discussion as brief as
possible, we give here only the main results of this model. Full details are
provided in the appendix.

There are three main contributions to the energy needed to create a
pseudospin texture in a LCR. Since in the ground state the in-plane
pseudospin component in a LCR is fully polarized along $S_{x}$, adding a
pseudospin texture has a tunnel energy cost when $t\neq 0$ because of the
interaction of the texture with the other channels. A second contribution
comes from the interlayer exchange interaction which is responsible for the
pseudospin stiffness $\rho _{s}$. As we mentioned above, the exchange
interaction favors pseudospin textures that vary slowly in space. A third
contribution must be considered in our model in order to get agreement with
the microscopic approach. It is the Coulomb interaction between different
portions of the soliton in a channel. This interaction favors large solitons.

If the coherent channels are oriented along $y$ and are considered as
effectively one-dimensional, then the energy cost to make a pseudospin
texture on top of the ground state where all pseudospins point in the $x$
direction in each channel is

\begin{equation}
\delta E=\int dy\left[ \frac{1}{2}\rho _{s}\left( \frac{\partial \varphi (y)%
}{\partial y}\right) ^{2}-T\left[ \cos \varphi (y)-1\right] \right] .
\label{3_6}
\end{equation}%
where $\varphi (y)$ in the azimuthal angle of the pseudospins. Eq. (\ref{3_6}%
) is valid if we ignore the third contribution mentionned above. The
parameters $\rho _{s}$ and $T$ are the \textit{effective} stiffness and
tunneling parameters. These parameters depend on the precise shape of the
LCR's as well as on the interaction between pseudospins of different
channels. In the appendix, we derive a microscopic expression for each of
these parameters in terms of the order parameters of the CSP. We show that
the effective stiffness is given by 
\begin{equation}
\rho _{s}=\frac{-1}{16\pi ^{2}\ell ^{2}}\left( \frac{e^{2}}{\kappa \ell }%
\right) \int dq_{x}\lvert \Omega (q_{x})\rvert ^{2}\,\,\left. \frac{%
d^{2}X_{R,L}(N,N;\mathbf{q})}{dq_{y}^{2}}\right\vert _{q_{y}\rightarrow 0},
\end{equation}%
where 
\begin{equation}
\Omega (q_{x})=\xi \sum_{G_{x}}\left\langle \rho
_{N}^{x}(G_{x})\right\rangle \frac{\sin \left[ \left( G_{x}-q_{x}\right) \xi
/4\right] }{\left( G_{x}-q_{x}\right) \xi /4},
\end{equation}%
is a form factor that takes into account the shape of the channel centered
at $x=0$. Also, $\xi $ is the interstripe distance in the CSP, $G_{x}=2\pi
n/\xi $ with $n=0,\pm 1,\pm 2,...$ and $\left\langle \rho
_{N}^{x}(G_{x})\right\rangle =\mathrm{Re}\left[ \left\langle \rho
_{N}^{R,L}(G_{x})\right\rangle \right] .$ If we define the parameter $%
\widetilde{G}_{x}=4\pi n/\xi $ and%
\begin{equation}
J_{\bot }\left( \mathbf{q}\right) =-X_{R,L}\left( N,N;\mathbf{q}\right) ,
\end{equation}%
then the parameter $T$ can be written as 
\begin{widetext}%
\begin{equation}
T=\frac{1}{2\pi \ell ^{2}}\left( \frac{e^{2}}{\kappa \ell }\right) \left[ 
\widetilde{t}\Omega (q_{x}=0)-\frac{1}{\xi }\sum_{\widetilde{G}_{x}}J_{\bot
}\left( \widetilde{G}_{x},0\right) \lvert \Omega (G_{x})\rvert ^{2}+\frac{1}{%
2}\frac{1}{L_{x}}\sum_{q_{x}}J_{\bot }\left( q_{x},0\right) \left\vert
\Omega \left( q_{x}\right) \right\vert ^{2}\right].  \label{3_1}
\end{equation}%
\end{widetext}The second and third terms in Eq. (\ref{3_1}) come from the
fact that, because of the pseudospin stiffness, there is an energy cost to
rotate the pseudospins in one channel when the pseudospins in the other
channels remain fixed in their ground state position. The contribution of
these two terms increases the effective tunneling strength $T$. Since the
energy cost to create a pseudospin soliton is given by $E_{s}=8\sqrt{\rho
_{s}T}$ we see that this second term keeps $E_{s}$ finite even when $%
\widetilde{t}=0$.

In this field-theoretic model, the energy to create an antisoliton is the
same as that needed to create a soliton and the charge gap is simply given
by%
\begin{equation}
\Delta =16\sqrt{\rho _{s}T}.  \label{gapfield}
\end{equation}%
From the energy functional of Eq. (\ref{3_6}), we get that the static
solution that minimizes the energy must satisfy the sine-Gordon equation 
\begin{equation}
\frac{\partial ^{2}\varphi (y)}{\partial y^{2}}=\frac{T}{\rho _{s}}\sin
\varphi (y).
\end{equation}%
The sine-Gordon (or kink) soliton is a solution of this equation. It is
given by%
\begin{equation}
\varphi (y)=4\tan ^{-1}\left[ e^{\sqrt[-]{\frac{T}{\rho _{s}}}y}\right] .
\label{pseudso}
\end{equation}%
The length of the soliton can be defined as%
\begin{equation}
L_{s}=\sqrt{\frac{\rho _{s}}{T}}.  \label{solitonsize}
\end{equation}

With the energy functional of Eq. (\ref{3_6}), we find numerically that both 
$\rho _{s}$ and $T$ decrease rapidly with $d$ but the size of the soliton $%
L_{s}$ decreases with increasing $d$. This behavior is opposite to what we
obtain in the microscopic calculation where the soliton size increases with $%
d$. As we mentionned above, it is necessary to include the Coulomb
interaction between different part of the solitons in order to get the
soliton size to increase with $d$. This leads to the term (full details are
given in the appendix) 
\begin{equation}
\delta E_{Coul}=\frac{\ell ^{2}}{32\pi ^{2}}\int dy\int dy^{\prime }\frac{%
d\varphi \left( y\right) }{dy}V_{\text{\emph{eff}}}\left( y-y^{\prime }\right) \frac{%
d\varphi \left( y^{\prime }\right) }{dy^{\prime }}
\end{equation}%
Inclusion of this term in in the energy functional introduces a nonlocal
non-linear term in the differential equation for the soliton and the
resulting equation is very difficult to solve. Following S. Ghosh and R.
Rajaraman\cite{ghosh} who use a similar procedure in their calculation of
the energy of CP$^{3}$ skyrmions in bilayers, we make the following
approximation. We insert a pseudospin texture $\varphi (y)=4\tan ^{-1}\left[
e^{-y/L_{s}^{\ast }}\right] $ into the total energy functional including the
Coulomb integral and evaluate is as a function of $L_{s}^{\ast }$. We then
minimize the total energy with respect to the length $L_{s}^{\ast }$ to
obtain the energy and length of the soliton. In this way, we find a soliton
length that increases with $d$ as in the microscopic approach. The procedure
is described in details in the appendix.

\section{The supercell microscopic Hartree-Fock method}

Let $\varepsilon _{CSP}$ be the energy \textit{per electron} in the CSP at $%
\nu =4N+1$ and magnetic field $B_{0}$ in units of $e^{2}/\kappa \ell _{0}$.
If the number of electrons is kept constant and the magnetic field is
decreased (to $B_{1}$) or increased (to $B_{2}$) such that the filling
factor becomes $\nu =4N\pm \widetilde{\nu }$, then a finite density $%
n_{qp}=\left\vert \widetilde{\nu }-1\right\vert /2\pi \ell _{1,2}^{2}$ of
quasiparticles (solitons for $\widetilde{\nu }>1$ and antisolitons for $%
\widetilde{\nu }<1$) are created in the CSP. At zero temperature, we expect
these quasiparticles to crystallize and to be localized in the LCR's of the
CSP. In the limit where only one quasiparticle is created ($\widetilde{\nu }%
\rightarrow 1$), we can define the quasiparticle energy as 
\begin{equation}
E_{qp}^{\pm }=\lim_{N_{qp}\rightarrow 1}\frac{\nu }{\left\vert \widetilde{%
\nu }-1\right\vert }\left[ \varepsilon _{SC}\left( \frac{e^{2}}{\kappa \ell
_{1,2}}\right) -\varepsilon _{CSP}\left( \frac{e^{2}}{\kappa \ell _{0}}%
\right) \right] ,  \label{4_1}
\end{equation}%
where $\varepsilon _{SC}$ is the energy per electron in the soliton crystal
(SC) in units of $e^{2}/\kappa \ell $ with $N_{qp}$ solitons and $%
E_{qp}^{+}\left( E_{qp}^{-}\right) $ is the energy to create one soliton
(antisoliton).

The quasiparticle energy defined in this way, with the number of electrons
kept constant, is refered to as the \textquotedblleft
proper\textquotedblright\ quasiparticle energy by Morf and Halperin\cite%
{MorfHalperin}. Other definitions are also possible. For example, the
\textquotedblleft gross\textquotedblright\ quasiparticle energies (or
chemical potentials) are defined by%
\begin{eqnarray}
\mu ^{+} &=&E\left( N_{e}=N_{\phi }+1\right) -E\left( N_{e}=N_{\phi }\right)
,  \label{4_2} \\
\mu ^{-} &=&E\left( N_{e}=N_{\phi }\right) -E\left( N_{e}=N_{\phi }-1\right)
,  \label{4_3}
\end{eqnarray}%
where $N_{\phi }$ is the degeneracy of the Landau levels at a magnetic field 
$B_{0}$ such that $\nu =4N+1.$ The energy $E\left( N_{\phi }\right) $ is the
total energy of the CSP, and $E\left( N_{\phi }\pm 1\right) $ is the total
energy of the CSP with one more (less) particle in the form of a soliton
(antisoliton). In this case, the magnetic field is kept constant while the
number of particles changes. At zero temperature, this is precisely the
definition of the chemical potentials at filling factors slightly above or
below $\nu =4N+1.$

The different definitions of the the quasiparticle energies lead to
different numerical values. As discussed by MacDonald and Girvin\cite%
{macdoquasip}, however, the numerical value of the gap, $\Delta ,$ is the
same for both definitions so that we can write 
\begin{equation}
\Delta =\mu ^{+}-\mu ^{-}=E_{qp}^{+}+E_{qp}^{-}.
\end{equation}

With the formalism described in Sec. II, we can easily compute the
Hartree-Fock energy of a crystal of solitons located in the coherent
channels of the bilayer. That is, we can compute $\varepsilon _{SC}$, find $%
E_{qp}^{\pm }$ and then the energy gap. However, there are several
difficulties with this method that we address in this paper. The first one
is that the limit $n_{qp}\rightarrow 1$ cannot be achieved numerically since
that would require infinite matrices in the equation of motion for the
single-particle Green's function. In this work, we have succeeded in
computing $\varepsilon _{SC}$ at filling as small as $\widetilde{\nu }=1\pm
0.02.$ The second difficulty is that, when a finite density of
quasiparticles is present, $\varepsilon _{SC}$ includes the interaction
energy between quasiparticles. This interaction energy must be computed and
removed from $\varepsilon _{SC}.$ A third difficulty is related to the size
of the solitons (antisolitons). In Sec. III, we saw that the soliton size
becomes very large when the tunneling energy $\widetilde{t}\rightarrow 0$ or
when $d$ is large. In this case the size of the soliton is not given by Eq. (%
\ref{solitonsize}) but is limited by the lattice constant of the soliton
crystal. The quasiparticle energy, then, cannot be computed reliably when
the tunneling term is too small or the interlayer distance too big.

We now describe in more details our evaluation of $E_{qp}^{\pm }.$ To avoid
computing numerically the energy of the antisoliton crystal as well as that
of the soliton crystal, we use the particle-hole symmetry of the Hamiltonian
around $\nu =4N+1$ to relate the energies of the two crystals with the same
filling of quasiparticles. We define state $0$ as the CSP at $\nu =4N+1$,
state $1$ as the soliton crystal at $\nu _{1}=4N+\widetilde{\nu }_{1}$ and
state $2$ as the crystal of antisolitons at $\nu _{2}=4N+\widetilde{\nu }%
_{2} $. The filling factors $\widetilde{\nu }_{2}=2-\widetilde{\nu }_{1}$ so
that the lattice constants $a_{1}$ and $a_{2}$ of the two crystals are
related by $\ell _{1}/a_{1}=\ell _{2}/a_{2}$. The Hartree-Fock energy 
\textit{per electron} of the three states\ are given by Eq. (\ref{2_2p})
which we rewrite here as 
\begin{equation}
\frac{E_{m}}{N_{e}}=\left[ \left( \frac{\widetilde{\nu }_{m}}{\nu _{m}}%
\right) \varepsilon _{m}\left( \widetilde{\nu }_{m}\right) +\frac{1}{\nu _{m}%
}\Lambda \left( \widetilde{\nu }_{m}\right) \right] \left( \frac{e^{2}}{%
\kappa \ell _{m}}\right) .  \label{4_4}
\end{equation}%
We have defined%
\begin{gather}
\varepsilon _{m}\left( \widetilde{\nu }_{m}\right) =-\frac{2\widetilde{t}}{%
\widetilde{\nu }_{m}}\mathrm{Re}\left[ \langle \rho _{N}^{R,L}(0)\rangle _{m}%
\right] \\
+\frac{1}{2\widetilde{\nu }_{m}}\sum_{i,j}\sum_{\mathbf{G}\neq
0}H_{i,j}\left( N,N,\mathbf{G}\right) \left\langle \rho _{N}^{i,i}\left( -%
\mathbf{G}\right) \right\rangle _{m}\left\langle \rho _{N}^{j,j}\left( 
\mathbf{G}\right) \right\rangle _{m}  \notag \\
-\frac{1}{2\widetilde{\nu }_{m}}\sum_{i,j}\sum_{\mathbf{G}}X_{i,j}\left( N,N,%
\mathbf{G}\right) \left\langle \rho _{N}^{i,j}\left( -\mathbf{G}\right)
\right\rangle _{m}\left\langle \rho _{N}^{j,i}\left( \mathbf{G}\right)
\right\rangle _{m},  \notag
\end{gather}%
\newline
which is the energy per electron \textit{in the partially filled level}. The
last term in Eq. (\ref{4_4}) is the interaction energy with the filled level
with%
\begin{equation}
\Lambda \left( \widetilde{\nu }_{i}\right) =-2\Lambda _{1}-\Lambda _{2}%
\widetilde{\nu }_{i},
\end{equation}%
where 
\begin{eqnarray}
\Lambda _{1} &=&\sum_{n<N}\sum_{n^{\prime }<N}X_{R,R}\left( n,n^{\prime
},0\right) , \\
\Lambda _{2} &=&\sum_{n<N}X_{i,i}\left( n,N,0\right) .
\end{eqnarray}

{}From Eqs. (\ref{4_2}) and (\ref{4_3}), it is easy to see that the
cyclotron and Zeeman energies do not contribute to the transport gap $\Delta 
$ and so can be ignored in Eq. (\ref{4_4}). This is also true of the filled
levels since their contribution to the quasiparticle energies are given by%
\begin{eqnarray}
\left( E_{qp}^{+}\right) _{f.l.} &=&\lim_{N_{qp}\rightarrow 1}\frac{\nu _{1}%
}{\left\vert \widetilde{\nu }_{1}-1\right\vert }\left( \frac{e^{2}}{\kappa
\ell _{1}}\right) \frac{1}{\nu _{1}}\Lambda \left( \widetilde{\nu }%
_{1}\right) \\
&&-\lim_{N_{qp}\rightarrow 1}\frac{\nu _{1}}{\left\vert \widetilde{\nu }%
_{1}-1\right\vert }\frac{1}{4N+1}\Lambda \left( 1\right) \left( \frac{e^{2}}{%
\kappa \ell _{0}}\right)  \notag \\
&=&\left( \frac{e^{2}}{\kappa \ell _{0}}\right) \left[ \frac{1}{2}\Lambda
_{2}+3\Lambda _{1}\right] ,  \notag
\end{eqnarray}%
and%
\begin{eqnarray}
\left( E_{qp}^{-}\right) _{f.l.} &=&\lim_{N_{qp}\rightarrow 1}\frac{\nu _{2}%
}{\left\vert \widetilde{\nu }_{2}-1\right\vert }\left( \frac{e^{2}}{\kappa
\ell _{2}}\right) \frac{1}{\nu _{2}}\Lambda \left( \widetilde{\nu }%
_{2}\right) \\
&&-\left( \frac{e^{2}}{\kappa \ell _{0}}\right) \lim_{N_{qp}\rightarrow 1}%
\frac{\nu _{2}}{\left\vert \widetilde{\nu }_{2}-1\right\vert }\frac{1}{4N+1}%
\Lambda \left( 1\right)  \notag \\
&=&-\left( \frac{e^{2}}{\kappa \ell _{0}}\right) \left[ \frac{1}{2}\Lambda
_{2}+3\Lambda _{1}\right] ,  \notag
\end{eqnarray}%
so that $\left( E_{qp}^{+}\right) _{f.l.}+\left( E_{qp}^{-}\right)
_{f.l.}=0. $ In deriving these two equations, we have used%
\begin{equation}
\left( \frac{e^{2}}{\kappa \ell _{1}}\right) =\left( \frac{e^{2}}{\kappa
\ell _{0}}\right) \sqrt{\frac{\nu _{0}}{\nu _{1}}}.  \label{4_6}
\end{equation}

From the electron-hole symmetry, we get 
\begin{equation}
\varepsilon _{2}=\left( \frac{\widetilde{\nu }_{1}}{2-\widetilde{\nu }_{1}}%
\right) \left[ \varepsilon _{1}+\left( \frac{\widetilde{\nu }_{1}-1}{%
\widetilde{\nu }_{1}}\right) X\left( 0\right) \right] ,  \label{4_5}
\end{equation}
where 
\begin{equation}
X\left( 0\right) =X_{R,R}\left( N,N,\mathbf{0}\right) .
\end{equation}
Note that Eq. (\ref{4_5}) is exact only in the limit where $%
N_{qp}\rightarrow 1$ because the inter-well Hartree and Fock interactions
contained in $\varepsilon _{m}$ depend on the ratio $d/\ell $ and we have $%
d/\ell _{1}\neq d/\ell _{2}$.

Combining all results, we have for the energy gap%
\begin{eqnarray}
\Delta &=&\lim_{\Delta \nu \rightarrow 0}\frac{1}{\Delta \nu }\widetilde{\nu 
}_{1}\left[ \sqrt{\frac{\nu _{0}}{\nu _{1}}}+\sqrt{\frac{\nu _{0}}{\nu _{2}}}%
\right] \varepsilon _{1}\left( \frac{e^{2}}{\kappa \ell _{0}}\right) \\
&&+\lim_{\Delta \nu \rightarrow 0}\frac{1}{\Delta \nu }\left[ \sqrt{\frac{%
\nu _{0}}{\nu _{2}}}X\left( 0\right) -2\widetilde{\varepsilon }_{CSP}\right]
\left( \frac{e^{2}}{\kappa \ell _{0}}\right) ,
\end{eqnarray}%
where we have defined%
\begin{equation}
\varepsilon _{CSP}=\frac{1}{4N+1}\widetilde{\varepsilon }_{CSP}.
\end{equation}%
Simplifying, we get finally%
\begin{equation}
\Delta =\lim_{\Delta \nu \rightarrow 0}\left[ 2\frac{\widetilde{\nu }_{1}}{%
\Delta \nu }\varepsilon _{1}-\frac{2}{\Delta \nu }\widetilde{\varepsilon }%
_{CSP}+X\left( 0\right) \right] \left( \frac{e^{2}}{\kappa \ell _{0}}\right)
.  \label{4_9}
\end{equation}%
We remark that the change in the magnetic length $\ell $ due to the change
in the magnetic field makes no contribution to the energy gap. We could have
ignored it in Eq. (\ref{4_4}). In fact, the gap defined using Eq. (\ref{4_1}%
) and taking $e^{2}/\kappa \ell _{i}=e^{2}/\kappa \ell _{0}$ is the
so-called neutral energy gap\cite{macdoquasip} and it is equal to the other
two gaps that we introduced in this section.

Eq. (\ref{4_9}) can also be written as%
\begin{equation}
\Delta =2E_{qp}^{+}+\left[ 2\varepsilon _{CSP}+X\left( 0\right) \right]
\left( \frac{e^{2}}{\kappa \ell _{0}}\right) .  \label{4_10}
\end{equation}%
In the lowest Landau level, the energy gap at $\nu =1$ is due to the
excitation of a bimeron-antibimeron pair and the energy per electron in the
UCS\ is $\varepsilon _{UCS}\left( d\right) =\left[ -\widetilde{t}-\frac{1}{4}%
\left[ X\left( 0\right) +\widetilde{X}_{d}\left( 0\right) \right] \right] $
where $\widetilde{X}\left( 0\right) =X_{R,L}\left( N,N,\mathbf{0}\right) $.
Eq. (\ref{4_10}) can then be written, for this special case, as%
\begin{equation}
\Delta =2E_{qp}^{+}+2\left[ \varepsilon _{UCS}\left( d\right) -\varepsilon
_{UCS}\left( d=0,t=0\right) \right] \left( \frac{e^{2}}{\kappa \ell _{0}}%
\right) ,  \label{4_11}
\end{equation}%
which is just the form we used in Ref. \onlinecite{breybimeron}.

\section{Interaction between quasiparticles}

With the simplifications introduced in the preceding section, the energy $%
\varepsilon _{SC}$ that enters Eq. (\ref{4_1}) and Eq. (\ref{4_9})\ is given
by 
\begin{gather}
\varepsilon _{SC}=-\frac{2\widetilde{t}}{\widetilde{\nu }}\mathrm{Re}\left[
\langle \rho ^{R,L}(0)\rangle \right]  \label{5_12} \\
+\frac{1}{2\widetilde{\nu }}\sum_{i,j}\sum_{\mathbf{G}\neq 0}H_{i,j}\left( 
\mathbf{G}\right) \left\langle \rho ^{i,i}\left( -\mathbf{G}\right)
\right\rangle \left\langle \rho ^{j,j}\left( \mathbf{G}\right) \right\rangle
\notag \\
-\frac{1}{2\widetilde{\nu }}\sum_{i,j}\sum_{\mathbf{G}}X_{i,j}\left( \mathbf{%
G}\right) \left\langle \rho ^{i,j}\left( -\mathbf{G}\right) \right\rangle
\left\langle \rho ^{j,i}\left( \mathbf{G}\right) \right\rangle ,  \notag
\end{gather}%
where, to simplify the notation, we have left implicit the index $N$ of the
Landau level. The soliton crystal is a superposition of a CSP with order
parameters $\left\{ \left\langle \alpha ^{i,j}\left( \mathbf{G}\right)
\right\rangle \right\} $ (computed at $\nu =4N+1$) and a pure soliton
crystal (PSC) with order parameters $\left\{ \left\langle \beta ^{i,j}\left( 
\mathbf{G}\right) \right\rangle \right\} $ such that%
\begin{equation}
\left\langle \rho ^{i,j}\left( \mathbf{G}\right) \right\rangle =\left\langle
\alpha ^{i,j}\left( \mathbf{G}\right) \right\rangle +\left\langle \beta
^{i,j}\left( \mathbf{G}\right) \right\rangle .  \label{5_11}
\end{equation}%
If we insert this decomposition into Eq. (\ref{5_12}), we find%
\begin{equation}
\varepsilon _{SC}=\varepsilon _{CSP}\left( \widetilde{\nu }\right)
+\varepsilon _{CSP-PSC}+\varepsilon _{PSC},
\end{equation}%
where%
\begin{gather}
\varepsilon _{CSP}\left( \widetilde{\nu }\right) =-\frac{2\widetilde{t}}{%
\widetilde{\nu }}\mathrm{Re}\left[ \langle \alpha ^{R,L}(0)\rangle \right] \\
+\frac{1}{2\widetilde{\nu }}\sum_{i,j}\sum_{\mathbf{G}\neq 0}H_{i,j}\left( 
\mathbf{G}\right) \left\langle \alpha ^{i,i}\left( -\mathbf{G}\right)
\right\rangle \left\langle \alpha ^{j,j}\left( \mathbf{G}\right)
\right\rangle  \notag \\
-\frac{1}{2\widetilde{\nu }}\sum_{i,j}\sum_{\mathbf{G}}X_{i,j}\left( \mathbf{%
G}\right) \left\langle \alpha ^{i,j}\left( -\mathbf{G}\right) \right\rangle
\left\langle \alpha ^{j,i}\left( \mathbf{G}\right) \right\rangle  \notag
\end{gather}%
is the energy per electron of the CSP (\textit{i.e.} $\varepsilon _{CSP}\left( 
\widetilde{\nu }\right) =\frac{1}{\widetilde{\nu }}\widetilde{\varepsilon }%
_{CSP}$), 
\begin{gather}
\varepsilon _{PSC}=-\frac{2\widetilde{t}}{\widetilde{\nu }}\mathrm{Re}\left[
\langle \beta ^{R,L}(0)\rangle \right]  \label{5_13} \\
+\frac{1}{2\widetilde{\nu }}\sum_{i,j}\sum_{\mathbf{G}\neq 0}H_{i,j}\left( 
\mathbf{G}\right) \left\langle \beta ^{i,i}\left( -\mathbf{G}\right)
\right\rangle \left\langle \beta ^{j,j}\left( \mathbf{G}\right) \right\rangle
\notag \\
-\frac{1}{2\widetilde{\nu }}\sum_{i,j}\sum_{\mathbf{G}}X_{i,j}\left( \mathbf{%
G}\right) \left\langle \beta ^{i,j}\left( -\mathbf{G}\right) \right\rangle
\left\langle \beta ^{j,i}\left( \mathbf{G}\right) \right\rangle  \notag
\end{gather}%
is the energy per electron of the PSC and 
\begin{gather}
\varepsilon _{CSP-PSC}= \\
+\frac{1}{\widetilde{\nu }}\sum_{i,j}\sum_{\mathbf{G}\neq 0}H_{i,j}\left( 
\mathbf{G}\right) \left\langle \alpha ^{i,i}\left( -\mathbf{G}\right)
\right\rangle \left\langle \beta ^{j,j}\left( \mathbf{G}\right) \right\rangle
\notag \\
-\frac{1}{\widetilde{\nu }}\sum_{i,j}\sum_{\mathbf{G}}X_{i,j}\left( \mathbf{G%
}\right) \left\langle \alpha ^{i,j}\left( -\mathbf{G}\right) \right\rangle
\left\langle \beta ^{j,i}\left( \mathbf{G}\right) \right\rangle  \notag
\end{gather}%
is the interaction energy (per electron) between the CSP\ and the PSC.

The contribution $\varepsilon _{PSC}$ causes problem because it contains not
only the energy to create the $N_{qp}$ solitons but also the interaction
energy between the solitons. This interaction energy goes away in the limit $%
\Delta \nu \rightarrow 0$. As we said, however, we cannot go to arbitrarily
small $\Delta \nu $ numerically because solving the equation of motion for
the single-particle Green's function then involves diagonalizing very large
matrices. We must then find a way to remove the interaction energy in $%
\varepsilon _{PSC}$. Two methods can be used. The first one is to replace $%
\varepsilon _{PSC}$ by $\varepsilon _{PSC}-\varepsilon _{int}$ where $%
\varepsilon _{int}$ is the Madelung energy of the crystal of charged
quasiparticles, assuming the quasiparticles to be point particles\cite%
{breybimeron}. We refer to this method as the \textquotedblleft
Madelung\textquotedblright\ method. In the limit $\Delta \nu \rightarrow 0$,
the quasiparticles are very far apart and, if they have an isotropic charge
distribution, it is a reasonable approximation. In the second method, which
we refer to as the \textquotedblleft form factor\textquotedblright\ method,
we completely replace $\varepsilon _{PSC}\left( \left\{ \left\langle \beta
^{i,j}\left( \mathbf{G}\right) \right\rangle \right\} \right) $ by the
energy $N_{qp}\varepsilon _{PSC}\left( \left\{ \left\langle \beta
_{qp}^{i,j}\left( \mathbf{q}\right) \right\rangle \right\} \right) $ where $%
\varepsilon _{PSC}\left( \left\{ \left\langle \beta _{qp}^{i,j}\left( 
\mathbf{q}\right) \right\rangle \right\} \right) $ is the energy per
electron of a \textquotedblleft crystal\textquotedblright\ of only one
quasiparticle. In the case of solitons, which are quite extended and highly
anisotropic objects it is necessary to use this second approach.

To evaluate $\varepsilon _{PSC}\left( \left\{ \left\langle \beta
_{qp}^{i,j}\left( \mathbf{q}\right) \right\rangle \right\} \right) $, we
make use of the fact that, when the quasiparticles are very far apart (limit 
$\widetilde{\nu }\rightarrow 1$) so that there is no overlap of the density
or spin texture due to different quasiparticles, then we may think of the
order parameters in real space as given by%
\begin{equation}
\left\langle \beta ^{i,j}\left( \mathbf{r}\right) \right\rangle =\sum_{%
\mathbf{R}}h_{i,j}\left( \mathbf{r-R}\right) ,
\end{equation}%
where $\mathbf{R}$ is a lattice site. We know that 
\begin{equation}
\left\langle \beta ^{i,j}\left( \mathbf{r}\right) \right\rangle =\frac{1}{V}%
\sum_{\mathbf{G}}\left\langle \beta ^{i,j}\left( \mathbf{G}\right)
\right\rangle e^{-i\mathbf{G}\cdot \mathbf{r}},
\end{equation}%
but it is not possible to get $h_{i,j}\left( \mathbf{r}\right) $ from this
equation. We must make an approximation. Since we work in the low-density
limit for the quasiparticles, it is a good approximation to assume that for
a \textquotedblleft crystal\textquotedblright\ of one quasiparticle %
\begin{equation}
\left\langle \beta ^{i,j}\left( \mathbf{r}\right) \right\rangle
_{qp}=\left\{ 
\begin{array}{ccc}
\frac{1}{V}\sum_{\mathbf{G}}\left\langle \beta ^{i,j}\left( \mathbf{G}%
\right) \right\rangle e^{-i\mathbf{G}\cdot \mathbf{r}}, & \mathrm{if} & 
\mathbf{r}\in v_{c} \\ 
0, & \mathrm{if} & \mathbf{r\notin }v_{c}%
\end{array}%
\right. ,  \label{4_8}
\end{equation}%
where $v_{c}$ is the volume of the unit cell centered at $\mathbf{r}=0.$
Fourier transforming Eq. (\ref{4_8}), we have 
\begin{eqnarray}
\left\langle \beta ^{i,j}\left( \mathbf{q}\right) \right\rangle _{qp}
&=&\int_{V}d\mathbf{r}e^{i\mathbf{q}\cdot \mathbf{r}}\left\langle \beta
^{i,j}\left( \mathbf{r}\right) \right\rangle _{qp} \\
&=&\frac{1}{N_{qp}}\sum_{\mathbf{G}}\left\langle \beta ^{i,j}\left( \mathbf{G%
}\right) \right\rangle \Lambda \left( \mathbf{q}-\mathbf{G}\right) \mathbf{,}
\notag
\end{eqnarray}%
where the form factor 
\begin{equation}
\Lambda \left( \mathbf{q}-\mathbf{G}\right) =\frac{1}{v_{c}}\int_{v_{c}}d%
\mathbf{r}e^{i\mathbf{q}\cdot \mathbf{r}}e^{-i\mathbf{G}\cdot \mathbf{r}},
\end{equation}%
depends on the shape of the unit cell of the soliton crystal.

It now remains to compute the Hartree-Fock energy corresponding to the
density and pseudospin textures given by the $\left\langle \beta
_{i,j}\left( \mathbf{q}\right) \right\rangle _{qp}$'s. The energy
is still given by an equation similar to Eq. (\ref{5_13}) where the
summation $\frac{1}{2\widetilde{\nu }}\sum_{\mathbf{G}}$ is now replaced by $%
\frac{1}{2\widetilde{\nu }}\sum_{\mathbf{q}}$. To go from the sum to the
integral, we use 
\begin{eqnarray}
\frac{1}{2\widetilde{\nu }}\sum_{\mathbf{q}}\left( \ldots \right)
&\rightarrow &\frac{S}{2\widetilde{\nu }}\int \frac{d\mathbf{q}}{\left( 2\pi
\right) ^{2}}\left( \ldots \right) \\
&\rightarrow &\frac{2\pi \widetilde{N}_{e}}{2\widetilde{\nu }^{2}}\int \frac{%
d\mathbf{q}\ell ^{2}}{\left( 2\pi \right) ^{2}}\left( \ldots \right) . 
\notag
\end{eqnarray}%
Aso, because $\left\langle \beta ^{i,j}\left( \mathbf{0}\right)
\right\rangle _{qp}\sim 1/N_{\varphi }$, we introduce a new field $\Theta
^{i,j}\left( \mathbf{q}\right) $ by the definition%
\begin{equation}
\Theta ^{i,j}\left( \mathbf{q}\right) =N_{\varphi }\left\langle \beta
^{i,j}\left( \mathbf{q}\right) \right\rangle _{qp}.
\end{equation}%
With this last definition, we have%
\begin{eqnarray}
&&N_{qp}\varepsilon _{PSC}\left( \left\{ \left\langle \beta
_{qp}^{i,j}\left( \mathbf{q}\right) \right\rangle \right\} \right) \\
&=&\frac{\pi \Delta \nu }{\widetilde{\nu }}\sum_{i,j}\int \frac{d\mathbf{q}%
\ell ^{2}}{\left( 2\pi \right) ^{2}}H_{i,j}\left( \mathbf{q}\right) \Theta
^{i,i}\left( -\mathbf{q}\right) \Theta ^{j,j}\left( \mathbf{q}\right)  \notag
\\
&&-\frac{\pi \Delta \nu }{\widetilde{\nu }}\sum_{i,j}\int \frac{d\mathbf{q}%
\ell ^{2}}{\left( 2\pi \right) ^{2}}X_{i,j}\left( \mathbf{q}\right) \Theta
^{i,j}\left( -\mathbf{q}\right) \Theta ^{j,i}\left( \mathbf{q}\right) . 
\notag
\end{eqnarray}

\begin{figure}[tbph]
\includegraphics[scale=1]{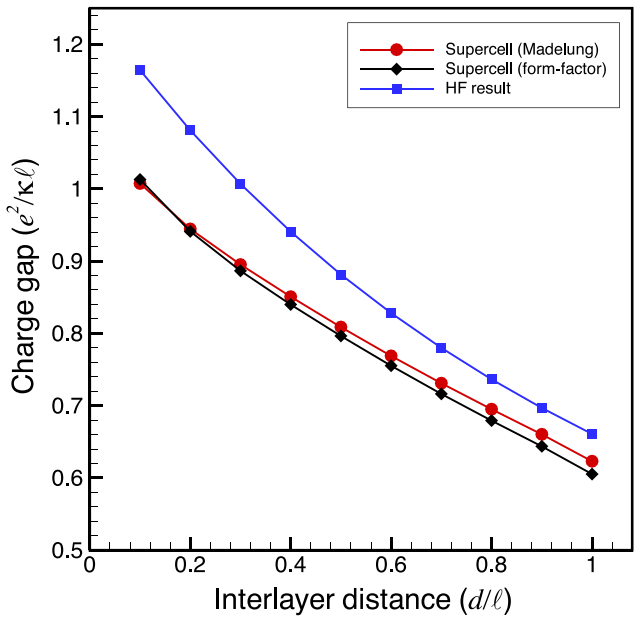}
\caption{The energy gap due to the excitation of a bimeron-antibimeron pair $%
\protect\nu =1$ computed using the form factor or the Madelung method and
compared with the Hartree-Fock energy gap to the excitation of an
electron-hole pair.}
\label{comparaison}
\end{figure}

As a test of our \textquotedblleft form factor\textquotedblright\ method, we
have computed the energy gap due to the creation of bimeron-antibimeron pairs at $%
\nu =1$ in the lowest Landau level $N=0.$ Figure \ref{comparaison} shows the
energy gap computed from a triangular lattice of bimerons at $\nu =1.02$ and 
$\widetilde{t}=0.0025.$ In this case, the Madelung and form factor methods
give identical results at small interlayer distances while the Madelung
method slighlty overestimates the energy gap at higher distances. The
difference between the two approches at large $d$ is due to the fact that
the charge density profile of the bimeron becomes more and more anisotropic
as $d$ increases. Also, the Coulomb interaction is stronger between point
particles than between extended particles so that the Madelung approach
overestimates the gap energy.

\begin{figure}[tbph]
\includegraphics[scale=1]{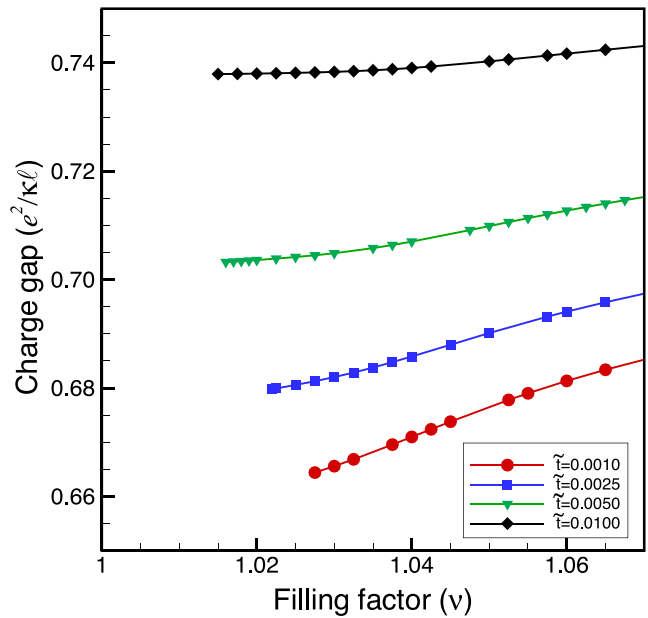}
\caption{Energy gap of the UCS at $\protect\nu =1$\ computed by the
supercell approach using the form factor method. The different curves are
for different values of the tunneling strength $\widetilde{t}.$ }
\label{gapmulti}
\end{figure}

To check the convergence of the supercell approach as the lattice constant
gets very large, we show in Fig. \ref{gapmulti} the energy gap of the UCS at 
$\nu =1$ computed at different values of $\nu $ from a crystal of bimerons.
The different curves in this figure are for different values of the
tunneling strength. The real gap of the system is, of course, defined for $%
\nu \rightarrow 1.$ We see that the gap converges more rapidly to its $\nu
\rightarrow 1$ value when the tunneling is stronger. This is understandable
since the size of a bimeron decreases when $\widetilde{t}$ increases and,
for sufficiently strong $\widetilde{t}$, this size is independent of the
lattice constant even at relatively high $\nu $. For smaller $\widetilde{t}$
the gap converges to its $\nu \rightarrow 1$ value, but only at lower
filling $\nu $. In the application of the supercell technique to the soliton
gap in the next section, we will use the form factor method to remove the
interaction energy. As we have just shown, this method is more appropriate
in the case where the quasiparticle is highly anisotropic in shape. 

\section{Numerical results}

\begin{figure}[tbph]
\includegraphics[scale=1]{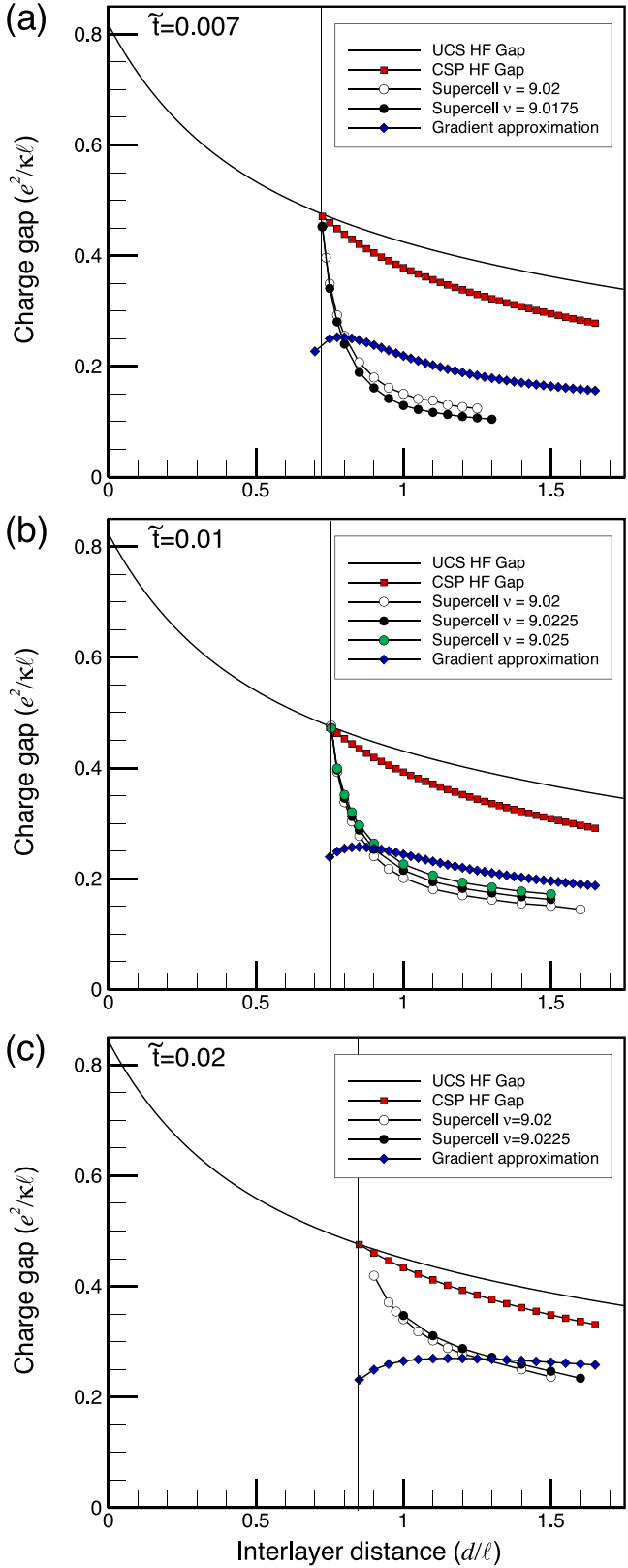}
\caption{Different energy gaps in the UCS and CSP calculated as a function
of the interlayer distance $d/\ell $ and for different values of the
tunneling parameter. For the supercell method, the gap is evaluated at
different filling factors to show the convergence of the results to the true
gap at $\protect\nu =9$. The gradient approximation refers to the
field-theoretic method.}
\label{gaps}
\end{figure}

We now discuss our numerical results for the energy gap of the CSP. Our
calculations are done in Landau level $N=2$ around $\nu =9$ using the form
factor method. Figures \ref{gaps}(a)-(c) contain our main results.
Differents gaps are plotted as a function of the interlayer distance for
tunnelings (a) $\widetilde{t}=0.007;$ (b) $\widetilde{t}=0.01$; and (c) $%
\widetilde{t}=0.02$. The filled line is $\Delta _{UCS},$ the energy needed
to create an ordinary electron-hole pair from the coherent liquid state at $%
\nu =9.$ At $\nu =9$, the liquid state is unstable for $d>d_{1}$ where the
coherent striped state is the ground state. The Hartree-Fock gap represented
by the curve with the filled squares is given by the energy to create an
electron-hole pair \textit{in a coherent channel} (see Fig. \ref%
{bandestructure} where this gap is defined). The other curves give the
energy gap calculated in the supercell method for different filling factors $%
\nu $ and the energy gap calculated with the field-theoretic approach
explained in the appendix. 

\begin{figure}[tbph]
\includegraphics[scale=1]{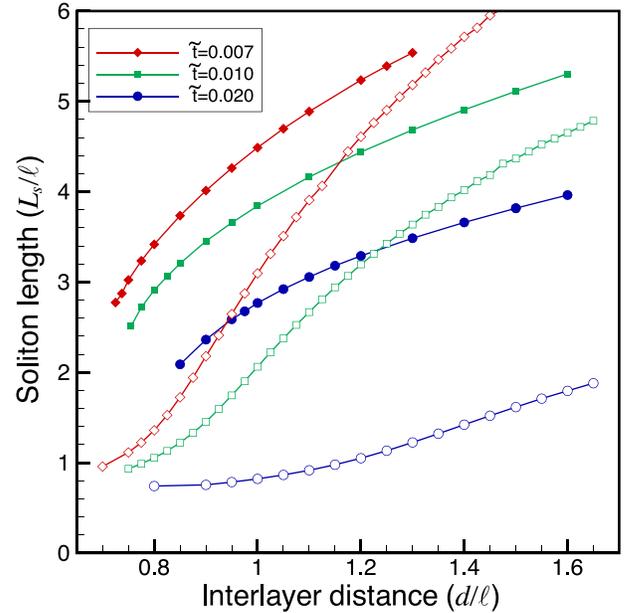}
\caption{Soliton size calculated with the supercell (filled symbols) and
field-theoretic (empty symbols) methods as a function of the interlayer
distance at $\protect\nu =9.02.$ In the supercell approach the size of the
soliton is found by fitting the $y$ dependence of the phase in a channel
with $\protect\varphi (y)=4\tan ^{-1}\left[ e^{-y/L_{s}}\right] .$ }
\label{tallei}
\end{figure}

{}From Fig. \ref{gaps}, it is clear that, in the CSP, the energy needed to
create a soliton-antisoliton pair is smaller than that needed to create an
electron-hole pair for typical experimental values of the tunneling
parameter $\widetilde{t}$. The transport gap is thus determined by the
creation of these topological excitations (as it was the case for skyrmions
in quantum Hall ferromagnet at $\nu =1$ or with bimerons in bilayer quantum
Hall systems).\cite{breybimeron} Figures \ref{gaps}(a)-(c) show a rapid
decrease of the energy gap near the transition between the coherent liquid
and the CSP that should be observable experimentally. The curves
corresponding to different filling factors show that the convergence of the
supercell method is quite good near the liquid-CSP transition but slow at
larger values of interlayer distances. This slow convergence is due to the
fact that the size of the soliton increases with interlayer distance as
shown in Fig. \ref{tallei} and the shape of the soliton is then restricted
by the lattice constant as we explained previously. As $d/\ell $ increases,
it becomes necessary to go to lower filling factors to achieve convergence,
something we cannot do numerically. In any case, the soliton gap is always
lower than the Hartree-Fock gap at higher values of $d/\ell $ since our
approach overestimates the energy gap. Increasing $\widetilde{t}$ decreases
the size of the solitons, however, so that it is possible to achieve better
convergence by increasing the value of the tunneling parameter $\widetilde{t}
$. This is seen by comparing Fig. \ref{gaps} (a), (b) and (c). Notice also
that, for smaller solitons, the soliton gap is closer to the Hartree-Fock
result, as expected. 

We also show in Fig. \ref{gaps} the gap calculated with the field-theoretic
method (see Eq. (\ref{gapfield})). This gap has the same qualitative
behavior with interlayer distance, except at small $d$ near the phase
transition. It is larger than the gap calculated in the microscopic approach.%
\textbf{\ }As we explain in the appendix, the field-theoretic result is
incorrect at small $d$ or large $\widetilde{t}$ (fig.  \ref{gaps}(c)) where
the stripes are not fully developped. At large $d$, we cannot say how
different the two gaps (macroscopic and field-theoretic) are because the gap
found in the microscopic approach has not yet converged at the lowest
filling factor we can achieve.

In the field-theoretic method, the soliton size, $L_{s}^{\ast },$ is
obtained by the procedure outlined in Sec. III . When the Coulomb
interaction between parts of the soliton is properly included, we find
numerically that $L_{s}^{\ast }$ increases with $d$ as in the supercell
calculation. Both approaches give the same trend for the soliton length. The
detailed behaviour with $d/\ell $ is quite different, however. Cearly, the
field-theoretic calculation does not capture all the subtleties of the
We recall that, as the
interlayer distance increases, the width of the LCR's becomes smaller. The
behavior of the soliton size may be understood as arising from the Coulomb
energy, which favors spreading the charge of the soliton. Our results are
plotted in Fig. \ref{tallei}. In this figure, we see that the supercell and
field-theoretic results do not match for large $\widetilde{t}$. This is
again due to the fact that the stripes are not fully formed at large $%
\widetilde{t}$ so that the expression of Eq. (\ref{pontry2}) for the
topological charge is not correct. As expected, Fig. \ref{tallei} shows that
the soliton size decreases with $\widetilde{t}$.

We have neglected quantum fluctuations in our calculation. These
fluctuations increases in importance as $d/\ell $ increases. They
renormalize the pseudospin stiffness and will probably also change the size
of the solitons and the quantitative values of the energy gaps. Inclusion of
these fluctuations is, however, beyond the scope of this paper.

\section{Conclusion}

We have computed the energy gap due to the creation of a soliton-antisoliton
pair in the linearly coherent channel of the coherent striped phase found in
higher Landau levels in a bilayer quantum Hall system. We have computed this
gap using a microscopic unrestricted Hartree-Fock approach as well as a
field-theoretic approach valid in the limit of slowly varying pseudospin
texture. With both methods, we find that the this energy gap is lower in
energy than the Hartree-Fock gap due to the creation of an electron-hole
pair in a coherent channel (a single spin flip) so that solitonic
excitations must play an important role in the transport properties of the
coherent striped phase.

\section{Acknowledgements}

This work was supported by a research grant (for R.C.) and graduate research
grants (for C. B. D.) both from the Natural Sciences and Engineering
Research Council of Canada (NSERC). H.A.F. acknowledges the support of NSF
through Grant No. DMR-0454699.

\appendix*

\section{Microscopic expressions for the parameters of the field-theoretic
model}

In this appendix we present the details of the derivation of the microscopic
expressions for the parameters $\rho _{s}$ and $T$ used in the
field-theoretic model of Sec. III. We drop the Landau level index $N$ here
since all order parameters are to evaluated in the partially filled level $N$%
. We begin by defining the pseudospin density operators 
\begin{align}
\rho (\mathbf{q})& =\rho ^{R,R}(\mathbf{q})+\rho ^{L,L}(\mathbf{q}), \\
\rho _{z}(\mathbf{q})& =\frac{1}{2}\left[ \rho ^{R,R}(\mathbf{q})-\rho
^{L,L}(\mathbf{q})\right] , \\
\rho _{x}(\mathbf{q})& =\frac{1}{2}\left[ \rho ^{R,L}(\mathbf{q})+\rho
^{L,R}(\mathbf{q})\right] , \\
\rho _{y}(\mathbf{q})& =\frac{1}{2i}\left[ \rho ^{R,L}(\mathbf{q})-\rho
^{L,R}(\mathbf{q})\right] .
\end{align}%
The total Hartree-Fock energy of the electrons in the partially filled level
for an unbiased bilayer can be written as%
\begin{equation}
E_{HF}=\varepsilon \left( \frac{e^{2}}{\kappa \ell }\right) ,
\end{equation}%
where 
\begin{eqnarray}
\varepsilon &=&-2N_{\phi }\widetilde{t}\left\langle \rho _{x}\left( 0\right)
\right\rangle  \label{a_10} \\
&&+\frac{1}{4}N_{\phi }\sum_{\mathbf{q}}\Upsilon \left( \mathbf{q}\right)
\left\langle \rho \left( -\mathbf{q}\right) \right\rangle \left\langle \rho
\left( \mathbf{q}\right) \right\rangle  \notag \\
&&+N_{\phi }\sum_{\mathbf{q}}J_{z}\left( \mathbf{q}\right) \left\langle \rho
_{z}\left( -\mathbf{q}\right) \right\rangle \left\langle \rho _{z}\left( 
\mathbf{q}\right) \right\rangle  \notag \\
&&+N_{\phi }\sum_{\mathbf{q}}J_{\bot }\left( \mathbf{q}\right) [\left\langle
\rho _{x}\left( -\mathbf{q}\right) \right\rangle \left\langle \rho
_{x}\left( \mathbf{q}\right) \right\rangle  \notag \\
&&\qquad +\left\langle \rho _{y}\left( -\mathbf{q}\right) \right\rangle
\left\langle \rho _{y}\left( \mathbf{q}\right) \right\rangle ].  \notag
\end{eqnarray}%
We have introduced the interactions 
\begin{equation}
J_{z}\left( \mathbf{q}\right) =H_{R,R}\left( \mathbf{q}\right)
-H_{R,L}\left( \mathbf{q}\right) -X_{R,R}\left( \mathbf{q}\right) ,
\end{equation}%
\begin{equation}
\Upsilon \left( \mathbf{q}\right) =H_{R,R}\left( \mathbf{q}\right)
+H_{R,L}\left( \mathbf{q}\right) -X_{R,R}\left( \mathbf{q}\right) ,
\label{a_11}
\end{equation}%
and 
\begin{equation}
J_{\bot }\left( \mathbf{q}\right) =-X_{R,L}\left( \mathbf{q}\right) .
\end{equation}%
In Eq. (\ref{a_10}), $H_{R,R}\left( 0\right) =H_{R,L}\left( 0\right) =0$
because of the interaction between the 2DEG and the positive background of
the donors.

We now introduce a unitless and unitary pseudospin field $S_{\alpha }(%
\mathbf{r})$, with $\alpha =x,y,z$ related to the guiding center density
operators in the pseudospin formalism by the relation 
\begin{equation}
S_{\alpha }(\mathbf{r})=4\pi \ell ^{2}N_{\phi }\left\langle \rho _{\alpha }(%
\mathbf{r})\right\rangle ,
\end{equation}%
and a projected\cite{note1} electron density by the relation 
\begin{equation}
n(\mathbf{r})=N_{\phi }\left\langle \rho (\mathbf{r})\right\rangle .
\end{equation}%
Using the definition of the pseudospin operators $S_{\alpha }({r})$ and
taking the Fourier transform of Eq. (\ref{a_10}), we have 
\begin{eqnarray}
\varepsilon &=&\frac{-\widetilde{t}}{2\pi \ell ^{2}}\int d\mathbf{r}S_{x}(%
\mathbf{r}) \\
&&+\frac{1}{8\pi \ell ^{2}}\int d\mathbf{r}\int d\mathbf{r}^{\prime
}J_{\perp }(\mathbf{r}-\mathbf{r}^{\prime })\mathbf{S}_{\perp }(\mathbf{r}%
)\cdot \mathbf{S}_{\perp }(\mathbf{r}^{\prime })  \notag \\
&&+\frac{1}{8\pi \ell ^{2}}\int d\mathbf{r}\int d\mathbf{r}^{\prime }S_{z}(%
\mathbf{r})J_{z}(\mathbf{r}-\mathbf{r}^{\prime })S_{z}(\mathbf{r}^{\prime })
\notag \\
&&+\frac{\pi \ell ^{2}}{2}\int d\mathbf{r}\int d\mathbf{r}^{\prime }n(%
\mathbf{r})\Upsilon (\mathbf{r}-\mathbf{r}^{\prime })n(\mathbf{r}^{\prime }).
\notag
\end{eqnarray}

Writing $S_{\alpha }(\mathbf{r})$ in spherical coordinates, it is easy to
describe the CSP ground state as 
\begin{align}
S_{x}(\mathbf{r})_{CSP}& =\sin \theta (x), \\
S_{y}(\mathbf{r})_{CSP}& =0, \\
S_{z}(\mathbf{r})_{CSP}& =\cos \theta (x),
\end{align}%
while the density $\left\langle n(\mathbf{r})\right\rangle =cst$ is uniform.
For a state where there is a spin texture only in the channel centered at $%
x=0$ (channel $0$) while the other channels remain in their CSP ground state
configuration (we recall that $\xi $ is the interstripe distance), we write%
\begin{equation}
S_{x}(\mathbf{r})=%
\begin{cases}
\sin \theta (x)\cos \varphi (y), & \text{if $|x|\leq \frac{\xi }{4}$}, \\ 
\sin \theta (x), & \text{if $|x|>\frac{\xi }{4}$},%
\end{cases}%
\end{equation}%
\begin{equation}
S_{y}(\mathbf{r})=%
\begin{cases}
\sin \theta (x)\sin \varphi (y), & \text{if $|x|\leq \frac{\xi }{4}$}, \\ 
0, & \text{if $|x|>\frac{\xi }{4},$}%
\end{cases}%
\end{equation}%
\begin{equation}
S_{z}(\mathbf{r})=S_{z}(\mathbf{r})_{CSP},
\end{equation}%
\begin{equation}
n(\mathbf{r})=n(\mathbf{r})_{CSP}+\delta n(\mathbf{r}).
\end{equation}%
In these equations, $\theta (x)$ is given by its value in the CSP. Defining 
\begin{equation}
J_{i,j}(y-y^{\prime })\equiv \int_{C_{i}}dx\int_{C_{j}}dx^{\prime }J_{\perp
}(\mathbf{r}-\mathbf{r^{\prime }})\sin \theta (x)\sin \theta (x^{\prime }),
\end{equation}%
where $C_{i}$ corresponds to the $i$-th channel of width $\xi /2$ centered at $x_i$ and $%
\int_{C_{i}}=\int_{x_i-\xi /4}^{x_i+\xi /4}$ , it is easy to show that the energy
difference between the this last state and the CSP ground state \textit{i.e.}
the energy to create one soliton in a channel is given by%
\begin{equation}
\begin{split}
\delta \varepsilon & =\frac{-\widetilde{t}}{2\pi \ell ^{2}}%
\int_{C_{0}}dx\sin \theta (x)\int dy\left[ \cos \varphi (y)-1\right] \\
+& \frac{1}{4\pi \ell ^{2}}\sum_{i\neq 0}\int dy\int dy^{\prime
}J_{i,0}(y-y^{\prime })\left[ \cos \varphi (y^{\prime })-1\right] \\
+& \frac{1}{8\pi \ell ^{2}}\int dy\int dy^{\prime }J_{0,0}(y-y^{\prime })%
\left[ \cos (\varphi (y)-\varphi (y^{\prime }))-1\right] \\
& +\frac{\pi \ell ^{2}}{2}\int d\mathbf{r}\int d\mathbf{r}^{\prime }\delta n(%
\mathbf{r})\Upsilon (\mathbf{r}-\mathbf{r}^{\prime })\delta n(\mathbf{r}%
^{\prime }) \\
& +\pi \ell ^{2}\int d\mathbf{r}\int d\mathbf{r}^{\prime }\delta n(\mathbf{r}%
)\Upsilon (\mathbf{r}-\mathbf{r}^{\prime })n(\mathbf{r}^{\prime })_{CSP}.
\end{split}
\label{a_2}
\end{equation}%
The first two terms contribute to the effective tunnelling term $T$ while
the third term is directly related to the pseudospin stiffness of the
system. The fourth term takes into account the Coulomb interaction between
different parts of the soliton and the last term is the interaction between
the charge of the soliton and that of the CSP. In an antisoliton, this fifth
contribution would have exactly the same value but with opposite sign so
that this last term does not contribute to the transport gap.

\subsection{Calculation of the pseudospin stiffness $\protect\rho _{s}$}

To extract the pseudospin stiffness from the third term of Eq. (\ref{a_2}),
we make a long-wavelength expansion of the $\cos (\varphi (y)-\varphi
(y^{\prime }))-1$ term. This expansion is possible if the pseudospin texture
varies slowly in comparison with $J_{0,0}(y)$. We get 
\begin{align}
& \frac{1}{8\pi \ell ^{2}}\int dy\int dy^{\prime }J_{0,0}(y-y^{\prime })%
\left[ \cos (\varphi (y)-\varphi (y^{\prime }))-1\right]  \notag \\
& =-\frac{1}{16\pi \ell ^{2}}\left[ \int dy^{\prime }\,\,y^{\prime
}{}^{2}J_{0,0}(y^{\prime })\right] \int dy\left( \frac{d\varphi (y)}{dy}%
\right) ^{2}.
\end{align}%
Comparing this last result with Eq. (\ref{3_6}), we see that 
\begin{equation}
\rho _{s}=-\frac{1}{8\pi \ell ^{2}}\int dy\,\,y^{2}J_{0,0}(y).
\end{equation}%
The pseudospin stiffness can be written, more explicitely as 
\begin{eqnarray}
\rho _{s} &=&-\frac{1}{8\pi \ell ^{2}}\int dy\,\,y^{2}\frac{1}{L_{x}L_{y}}%
\sum_{\mathbf{q}}J_{\perp }(\mathbf{q})e^{iq_{y}y} \\
&&\times \int_{C_{0}}dx\int_{C_{0}}dx^{\prime }\sin \theta (x)\sin \theta
(x^{\prime })e^{iq_{x}(x-x^{\prime })},  \notag
\end{eqnarray}%
with $L_{x}$ and $L_{y}$ the length and width of the sample. This allows the
integrals over $x$ and $x^{\prime }$ to be totally decoupled. In fact,
defining the form factor%
\begin{eqnarray}
\Omega (q_{x}) &=&\int_{C_{0}}dx\sin \theta (x)e^{iq_{x}x} \\
&=&\xi \sum_{G_{x}}\langle \rho _{x}(G_{x})\rangle \frac{\sin \left[
(G_{x}-q_{x})\xi /4\right] }{(G_{x}-q_{x})\xi /4},  \notag
\end{eqnarray}%
we can write%
\begin{equation}
\rho _{s}=\frac{1}{16\pi ^{2}\ell ^{2}}\int dq_{x}\lvert \Omega
(q_{x})\rvert ^{2}\,\,\left. \frac{d^{2}J_{\perp }(\mathbf{q})}{dq_{y}^{2}}%
\right\vert _{q_{y}\rightarrow 0}.
\end{equation}%
The form factor $\Omega (q_{x})$ takes into account the influence of the
shape of the charge modulation along the $x$ axis in the CSP phase on the
effective pseudospin stiffness in the one dimensional sine-Gordon model.

\subsection{Calculation of the tunneling parameter $T$}

The effective tunnel coupling $T$ can be extracted from the first two terms
of Eq. (\ref{a_2}). The first term renormalizes the tunnel coupling in the
1D effective theory, taking into account that interlayer coherence exists
only in the LCR's. This first term is simply

\begin{equation}
\frac{-\widetilde{t}}{2\pi \ell ^{2}}\Omega (0)\int dy\left[ \cos \varphi
(y)-1\right] .
\end{equation}

The second contribution to the effective tunnel coupling comes from the
exchange energy between channel $0$ (where a pseudospin texture was created)
and the other channels. In these other channels, the in-plane pseudospin
component is totally polarized along the $\mathbf{x}$ direction and the
exchange interaction between channel $i$ and channel $0$ favors a
configuration in channel $0$ where the pseudospin is also polarized along $+%
\mathbf{x}$, just like the simple tunnel coupling $\widetilde{t}$. In other
words, there is an energy cost, even in the absence of tunneling, to make a
rotation of the pseudospins in one channel because of the interaction with
the pseudospins in the other channels.

It is possible to extract a simple form for this coupling from the second
term of Eq. (\ref{a_2}) since 
\begin{eqnarray}
\sum_{i\neq 0}\int dyJ_{i,0}(y) &=&\frac{1}{L_{x}}\sum_{i\neq
0}\sum_{q_{x}}J_{\perp }(q_{x},0)|\Omega (q_{x})|^{2}e^{iq_{x}(x_{i}-x_{0})}
\notag \\
&=&\frac{1}{L_{x}}\sum_{i}\sum_{q_{x}}J_{\perp }(q_{x},0)|\Omega
(q_{x})|^{2}e^{iq_{x}(x_{i}-x_{0})}  \notag \\
&&-\frac{1}{L_{x}}\sum_{q_{x}}J_{\perp }(q_{x},0)|\Omega (q_{x})|^{2},
\end{eqnarray}%
with $x_{n}-x_{0}=n\xi /2$ the center-to-center distance between channels $n$
and $0$. Because there is a sum over the channels, the sum on the
wave-vectors $q_{x}$ reduces to a sum over the reciprocal lattice vectors of
a 1D lattice of lattice constant $\xi /2$, noted $\widetilde{G}_{x}$, and 
\begin{equation}
\begin{split}
& \frac{1}{2}\sum_{i\neq 0}\int dy\int dy^{\prime }J_{i,0}(y-y^{\prime })%
\left[ \cos \varphi (y^{\prime })-1\right] \\
& =\frac{1}{\xi }\sum_{\widetilde{G}_{x}}J_{\perp }(\widetilde{G}%
_{x},0)|\Omega (\widetilde{G}_{x})|^{2}-\frac{1}{2}\frac{1}{L_{x}}%
\sum_{q_{x}}J_{\perp }(q_{x},0)|\Omega (q_{x})|^{2}.
\end{split}%
\end{equation}%
Combining the two terms, we find 
\begin{eqnarray}
T &=&\frac{1}{2\pi \ell ^{2}}\left[ \Omega (0)\widetilde{t}-\frac{1}{\xi }%
\sum_{\widetilde{G}_{x}}J_{\perp }(\widetilde{G}_{x})\left\vert \Omega
\left( \widetilde{G}_{x}\right) \right\vert ^{2}\right. \\
&&\left. +\frac{1}{2}\frac{1}{L_{x}}\sum_{q_{x}}J_{\bot }\left(
q_{x},0\right) \left\vert \Omega \left( q_{x}\right) \right\vert ^{2}\right]
.  \notag
\end{eqnarray}

\subsection{Sine-Gordon soliton and the Coulomb energy}

If we combine the tunneling and exchange terms, we find that the energy cost
to make one soliton localized in a channel of the CSP\ is given by Eq. (\ref%
{3_6}). As we mentionned in Sec. III, the static solution that minimizes
this energy functional is the sine-Gordon (or kink) soliton $\varphi
(y)=4\tan ^{-1}\left[ e^{\sqrt[-]{\frac{T}{\rho _{s}}}y}\right] .$

We now add to Eq. (\ref{3_6}) the Coulomb interaction energy between
different parts of the soliton 
\begin{equation}
\delta E_{Coul}=\frac{\pi \ell ^{2}}{2}\int d\mathbf{r}\int d\mathbf{r}%
^{\prime }\delta n\left( \mathbf{r}\right) \Upsilon \left( \mathbf{r}-%
\mathbf{r}^{\prime }\right) \delta n\left( \mathbf{r}^{\prime }\right) .
\label{coulomb}
\end{equation}%
To relate $\delta n(\mathbf{r}^{\prime })$ to the angles $\theta $ and $%
\varphi $, we use the definition of the topological charge density given in
Eq. (\ref{pontry}). We assume that, in the one-soliton state, only $\varphi
\left( y\right) $ changes along a channel and that $\theta \left( \mathbf{r}%
\right) $ is given by its value in the CSP. We have 
\begin{eqnarray}
\delta n\left( \mathbf{r}\right) &=&-\frac{1}{4\pi }\nabla \varphi \left( 
\mathbf{r}\right) \times \left( \nabla \cos \theta \left( \mathbf{r}\right)
\right) \cdot \widehat{\mathbf{z}}  \label{pontry2} \\
&=&\frac{1}{4\pi }\frac{d\varphi \left( y\right) }{dy}\frac{d}{dx}\cos
\theta \left( x\right) .  \notag
\end{eqnarray}

At this point, we must remark that if we use the sine-Gordon solution in
Eq.~(\ref{pontry2}) and integrate the projected density $\delta n\left( 
\mathbf{r}\right) $ in a channel, we find $\int_{-\xi /4}^{+\xi
/4}dx\int_{-\infty }^{+\infty }dy\delta n\left( \mathbf{r}\right) =1$ only
if $\cos \theta \left( x\right) $ varies from $-1$ to $+1$ in the channel 
\textit{i.e.} only in the limit or large interlayer distances where the
stripes are fully developped. In consequence, we do not expect our
field-theoretic model to be valid near the transition between the UCS and
the CSP.

We insert Eq. (\ref{pontry2}) into Eq. (\ref{coulomb}), and define the form
factor (for a channel centered at $x=0$) 
\begin{eqnarray}
A\left( q_{x}\right) &=&\int_{C_{0}}dxe^{-iq_{x}x}\frac{d}{dx}\cos \theta
\left( x\right) \\
&=&i\xi \sum_{G_{x}}\left\langle \rho _{z}\left( G_{x}\right) \right\rangle
G_{x}\frac{\sin \left( q_{x}-G_{x}\right) \xi /4}{\left( q_{x}-G_{x}\right)
\xi /4},  \notag
\end{eqnarray}%
and the effective interaction $V_\text{\emph{eff}}\left( y-y^{\prime
}\right) $ in a channel%
\begin{equation}
V_{\text{\emph{eff}}}\left( y-y^{\prime }\right) =\frac{1}{S}\sum_{\mathbf{q}%
}\left\vert A\left( q_{x}\right) \right\vert ^{2}\Upsilon \left( \mathbf{q}%
\right) e^{iq_{y}\left( y-y^{\prime }\right) }.
\end{equation}%
We then find for the Coulomb interaction%
\begin{equation}
\delta E_{\text{\emph{Coul}}}=\frac{\ell ^{2}}{32\pi ^{2}}\int dy\int
dy^{\prime }\frac{d\varphi \left( y\right) }{dy}V_{\text{\emph{eff}}}\left(
y-y^{\prime }\right) \frac{d\varphi \left( y^{\prime }\right) }{dy^{\prime }}%
.
\end{equation}%
If we add the contribution $\delta E_{\text{\emph{Coul}}}$ to Eq. (\ref{3_6}%
)\ and minimize the energy with respect to $\varphi \left( y\right) $, we
find that it introduces a nonlocal term to the sine-Gordon equation. The
resulting equation is then very difficult to solve. To get an approximation
for the Coulomb energy, we decided to proceed in the following way. We take,
as a trial solution, the kink soliton 
\begin{equation}
\varphi (y)=4\tan ^{-1}\left[ e^{-y/L_{s}^{\ast }}\right] ,
\end{equation}%
where $L_{s}^{\ast }$ is the width of the soliton. The Coulomb energy is then%
\begin{equation}
\delta E_{\text{\emph{Coul}}}\left( L_{s}^{\ast }\right) =\frac{\pi \ell ^{2}%
}{32\pi ^{2}}\int d\mathbf{q}\left\vert A\left( q_{x}\right) \right\vert
^{2}\Upsilon \left( q\right) \text{sech} ^{2}\left( \frac{\pi
q_{y}L_{s}^{\ast }}{2}\right) .
\end{equation}%
The total energy for the soliton is%
\begin{equation}
E=4\frac{\rho _{s}}{L_{s}^{\ast }}+4TL_{s}^{\ast }+\delta E_{\text{\emph{Coul%
}}}\left( L_{s}^{\ast }\right) .
\end{equation}%
We find $L_{s}^{\ast }$ by minimizing numerically the total energy $E$. In
our numerical calculation, we use $\Upsilon \left( \mathbf{q}\right)
=H_{N}\left( \mathbf{q}\right) $ instead of Eq. (\ref{a_11}). This is also
the interaction considered in similar calculations\cite{macdobible},\cite%
{rajaraman}. The use of Eq. (\ref{a_11}) leads to non-physical results.


\begin{thebibliography}{99}
\bibitem{note1} The relation between the true density and the guiding center
density is given by $n(\mathbf{q})=N_{\phi }\rho (\mathbf{q})F(\mathbf{q})$
where $F(\mathbf{q})$ is a form factor that depends on the Landau level
index. In this work we use call $n(\mathbf{q})=N_{\phi }\rho (\mathbf{q})$
the projected density.
\end{thebibliography}
\end{document}